\documentclass{optica-article}

\journal{opticajournal} 

\articletype{Research Article}
\newcommand{\hlio}{H$_3$LiIr$_2$O$_6$} 
\newcommand{\Li}{$\alpha$-Li$_2$IrO$_3$}
\newcommand{\Na}{Na$_2$IrO$_3$}

\usepackage{lineno}
\usepackage{hyperref}

\newcommand{\OPc}[2]{\hat{#1}_{#2}^{\dag}}
\newcommand{\OP}[2]{\hat{#1}_{#2}^{\vphantom{\dag}}}
\newcommand{\CD}[1]{\OPc{c}{#1}}
\newcommand{\C}[1]{\OP{c}{#1}}
\newcommand{\ND}[1]{\hat{n}_{#1}}

\definecolor{cblue}{RGB}{19,107,192}
\hypersetup{colorlinks=true,linkcolor=cblue,citecolor=cblue,urlcolor=cblue}

\begin{document}

\title{Signatures of Floquet Engineering in the proximal Kitaev Quantum Spin Liquid \hlio{} by tr-RIXS}

\author{Jungho Kim,\authormark{1} Tae-Kyu Choi,\authormark{2} Edward Mercer,\authormark{3,4} Liam T. Schmidt,\authormark{3,4} Jaeku Park,\authormark{2} Sang-Youn Park,\authormark{2} Dogeun Jang,\authormark{2} Seo Hyoung Chang,\authormark{5} Ayman Said,\authormark{1} Sae Hwan Chun,\authormark{2} Kyeong Jun Lee,\authormark{5} Sang Wook Lee,\authormark{6} Hyunjeong Jeong,\authormark{6} Hyeonhui Jeong,\authormark{6} Chanhyeon Lee,\authormark{7} Kwang-Yong Choi, \authormark{7} Faranak Bahrami, \authormark{8} Fazel Tafti,\authormark{8} Martin Claassen,\authormark{9,10} and Alberto de la Torre\authormark{3,4,*}}

\address{\authormark{1}Advanced Photon Source, Argonne National Laboratory, Argonne, IL, USA}
\address{\authormark{2}XFEL Division, Pohang Accelerator Laboratory, POSTECH, Pohang, Gyeongbuk 37673, Republic of Korea}
\address{\authormark{3}Department of Physics, Northeastern University, Boston, MA, 02115, USA}
\address{\authormark{4}Quantum Materials and Sensing Institute, Northeastern University, Burlington, MA, 01803 USA}
\address{\authormark{5}Department of Physics, Chung-Ang University, Seoul 06974, Republic of Korea}
\address{\authormark{6}Department of Physics, Ewha Womans University, Seoul 03760, Republic of Korea}
\address{\authormark{7}Department of Physics, Sungkyunkwan University, Suwon 16419, Republic of Korea}
\address{\authormark{8}Department of Physics, Boston College, Chestnut Hill, MA, 02467, USA}
\address{\authormark{9}Department of Physics and Astronomy, University of Pennsylvania, Philadelphia, Pennsylvania 19104, USA}
\address{\authormark{10}Center for Computational Quantum Physics, Flatiron Institute, 162 5th Ave, New York, NY 10010}

\email{\authormark{*}a.delatorreduran@northeastern.edu} 


\begin{abstract*} We present the first circularly polarized Floquet engineering time-resolved Resonant Inelastic X-ray Scattering (tr-RIXS) experiment in \hlio{}, an iridium-based Kitaev system. Our calculations and experimental results are consistent with the modification of the low energy magnetic excitations in \hlio{} only during illumination by the laser pulse, consistent with the Floquet engineering of the exchange interactions. However, the penetration length mismatch between the X-ray probe and laser pump and the intrinsic complexity of Kitaev magnets prevented us from unequivocally extracting towards which ground \hlio{} was driven. We outline possible solutions to these challenges for Floquet stabilization and observation of the Kitaev Quantum Spin Liquid limit by RIXS.

\end{abstract*}



\section{Introduction}

\begin{figure}
    \centering
    \includegraphics[width=1\linewidth]{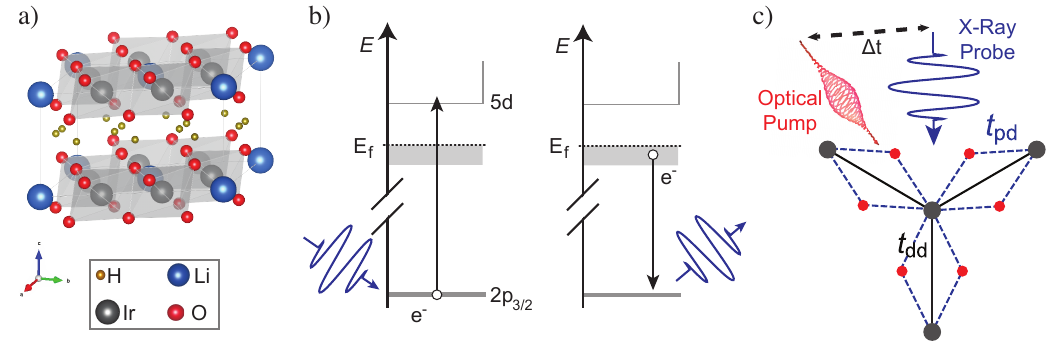}
    \caption{Tr-Floquet-RIXS in \hlio{}. a) Crystal structure of \hlio{}, illustrating the layered arrangement of the IrO$_{6}$ octahedra (gray) and interlayer $H$ ions. b) Schematic of the electronic transitions involved in the RIXS process at the Ir $L_{3}$ edge. Left: A resonant x-ray excites a $2p_{3/2}$ core electron into an unoccupied $5d$ state above the Fermi level, creating an electron-hole pair. Right: An electron below the Fermi level decays, annihilating the core-hole, emitting a photon and returning to the original state. c) Schematic of the tr-Floquet-RIXS approach. The circularly polarized optical pump will induce non-equilibrium dynamics, and then, after a delay ($\Delta t$), an x-ray probe is used to measure the system's response via RIXS. These inter-orbital transitions provide insight into excited-state dynamics and interactions within the IrO$_{6}$ lattice.}
    \label{fig:intro}
\end{figure}

The development of scalable topological quantum computing promises to address a major challenge in the field: the loss of quantum information due to thermal noise and decoherence \cite{SCHLOSSHAUER20191,onizhuk2024decoherencesolidstatespinqubits}. In a topological qubit, quantum information is encoded in the topological protected properties of Majorana zero modes, fractionalized excitations with non-Abelian statistics, making it robust to local distortions and fluctuations \cite{Universal_TQC,Easttom2024}. While experimental demonstration and control of Majorana fermions in superconductor-semiconductor nanowire devices remain elusive \cite{Cao2023}, the so-called Kitaev Quantum Spin Liquid (KQSL) systems might offer an alternative solid-state platform to access anyon statistics \cite{KITAEV20062,kitaev2009topologicalphasesquantumcomputation}. KQSL refers to exactly solvable quantum spin liquid state \cite{savary_quantum_2017,broholm_quantum_2020} emerging in a highly-frustrated set of spin-1/2 on a two-dimensional honeycomb lattice interacting via bond directional Ising-like interactions that hosts fractionalized Majorana and gauge flux excitations \cite{KITAEV20062}. Moreover, the KQSL was predicted to be realizable in strong spin-orbit coupling quantum materials hosting honeycomb planes with edge-sharing transition metal-ligand octahedra \cite{Rau_review,PhysRevLett.102.017205}, opening a new avenue for topological quantum computing.

However, most Kitaev candidate materials display magnetically ordered ground states \cite{Takagi2019}. Additional efforts to access the Kitaev limit via external fields, such as hydrostatic pressure and strong magnetic fields \cite{Stahl2024,Kasahara2018} or atomic substitution \cite{molecules27030871} have been prevented by the presence of a dimerization instability in the phase diagram of Kitaev magnets  and the resilience of the long-range ordered magnetic state to external tuning knobs fields \cite{PhysRevB.97.020104}. \hlio{} stands as the singular known candidate that lacks long-range order down to low temperature and exhibits all the characteristic physical properties associated with quantum spin liquids \cite{kitagawa_spinorbital-entangled_2018,geirhos_quantum_2020,Pei_Raman_2020}. Moreover, a recent momentum, energy, and temperature dependence Resonant Inelastic X-ray Scattering (RIXS) experiment at the Ir $L{_3}$ absorption-edge \cite{delaTorre2023} uncovered a broad continuum of magnetic excitations centered at $E = 25$~meV with a high energy tail extending up to $E = 170$~meV. This observation aligns with expectations for dominant ferromagnetic Kitaev interactions, $|K| = 25$~meV. Similar continuum of magnetic excitations were observed in inelastic neutron scattering measurement in D$_3$IrLi$_2$O$_6$ \cite{halloran2024}. However, the lack of translation symmetry of the magnetic continuum \cite{delaTorre2023,bette_solution_2017}, the divergence of the specific heat at low temperatures, the existence of non-vanishing contributions to the NMR response \cite{kitagawa_spinorbital-entangled_2018} and the time-field scaling of longitudinal field \textmu{}SR \cite{yang2024} indicate the presence of low energy excitations in \hlio{} due to a departure from the pure KQSL limit. As such, \hlio{} has been interpreted as displaying a dynamically fluctuating ground state near bond-disordered versions of the KQSL \cite{knolle_bond-disordered_2019}. Thus, the experimental realization of the KQSL remains elusive despite concerted synthesis and pressure engineering efforts.

An emerging approach to controlling quantum materials is to deploy ultrafast light-matter interaction to access phases of matter unstable in equilibrium \cite{Colloquium}. In this context, proposals to modify exchange interactions in magnetic Mott insulators by dressing the electron hopping by the oscillating laser electric field in the minimal coupling limit (Floquet description) have attracted much attention \cite{Mentink2015,claassen2017dynamical,PhysRevB.96.014406,Kennes2018,KennesPRB2019}. In this regime, if absorption resonances are avoided the charge subsystem remains unchanged, while a non-thermal regime can be reached in which local moments interact via renormalized magnetic interactions. When deployed to Kitaev magnets, Floquet engineering with a circularly polarized laser pulse has been numerically shown to enable the controllable and independent tuning of magnetic exchange interactions beyond Heisenberg exchange \cite{PhysRevB.103.L100408,Claassen_RuCl3,Kumar2022}. Thus, ultrafast light-matter interaction can be used to traverse the complex phase diagram of Kitaev magnets \cite{PhysRevLett.112.077204} as a function of pump laser frequency and fluence towards the transient KQSL limit. 

Here, we employ time-resolved RIXS (tr-RIXS) at the Pohang Accelerator Laboratory X-ray Free-Electron Laser (PAL-XFEL) facility in combination with circularly polarized strong 1900 nm laser pulses ($F = 97$ mJ/cm$^2$) to study the Floquet Engineering of magnetic exchange interactions in \hlio{}. tr-RIXS is a uniquely suited technique to access transient changes to the magnetic spectrum of quantum materials with sub-ps resolution. Our data, emerging from the first Floquet tr-RIXS experiment on a Kitaev magnet, are suggestive of an increase of coherence of the magnetic excitations only during pump illumination. On the other hand, calculations of the variation of the magnetic exchange interactions within the Floquet description as a function of laser pump frequency and fluence predict an evolution of \hlio{} toward the KQSL limit at the sample surface but less than $1\%$ change when average over the proven sample volume. This dichotomy exemplifies the intrinsic complexity of Kitaev magnets and of hard X-ray probe - laser pump experiments. We conclude with a discussion of how this pioneer experiment demonstrating transient changes to the magnetic excitations of a topical quantum material  opens new pathways to transiently access the elusive control of the KQSL phase.

\section{Materials and Methods}
\subsection{Sample growth and characterization}

The samples are from the same batch as in Ref. \cite{delaTorre2023}. Precursor single crystals of \Li{} were grown as described elsewhere \cite{Freund2016}. $40 \times 40 \mu$m single crystals of \hlio{} were grown via a topotactic exchange by placing \Li{} inside an autoclave filled with sulfuric acid for 48 hours \cite{bette_solution_2017}. The replacement of the inter-honeycomb layer Li with H leads to a modification of the intra-layer Ir-O-Ir bond angles and of the superexchange pathways. The proximal KQSL state was characterized by powder x-ray diffraction, specific heat, magnetization measurements \cite{molecules27030871}, and static RIXS. Before the tr-RIXS experiment, samples were aligned with by synchrotron XRD at the IVU - Undulator Beamline of the Canadian Light Source (CLS). For the tr-RIXS experiment, the scattering plane was set along $[H0L]$.

\subsection{tr-RIXS setup at the PAL-XFEL}

\begin{figure}
    \centering
    \includegraphics[width=1\linewidth]{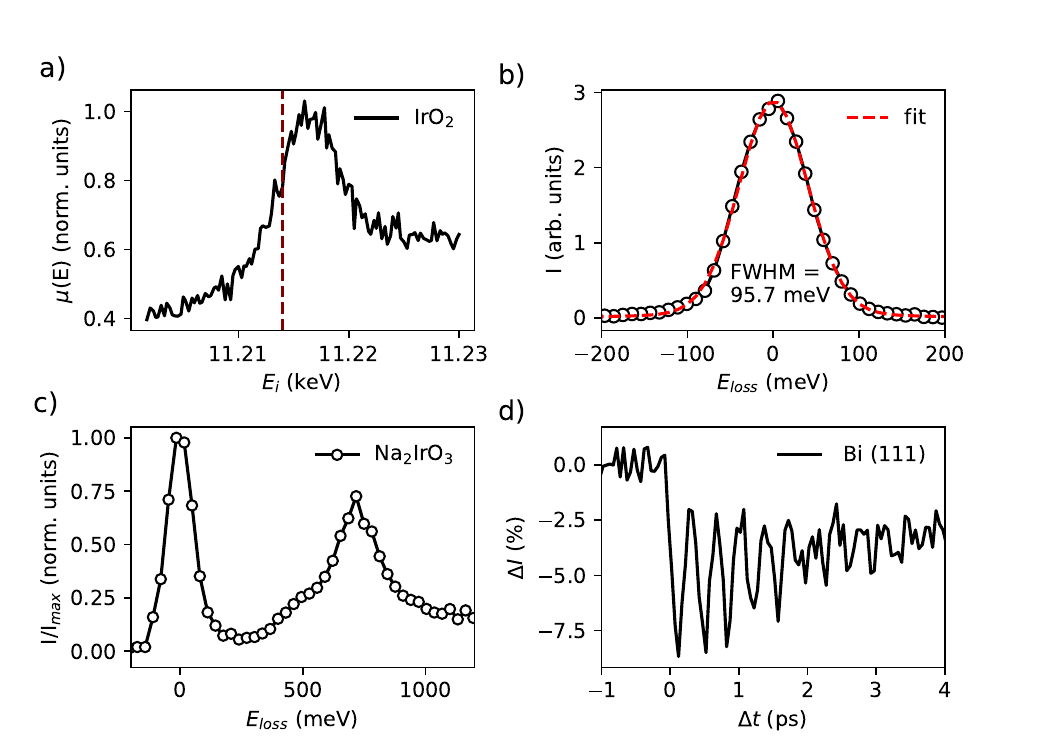}
    \caption{Characterization of the tr-RIXS experimental setup at the PAL-XFEL. a) X-ray absorption spectrum of IrO$_{2}$ powder sample measured in total fluorescence yield at room temperature. The vertical dashed line indicates the main Ir $L_3$ resonance at $E_{i} = 11214$~eV. b) Elastic scattering from a Scotch tape (circular markers) and fit to a Pseudo-Voigt profile (red dashed line) with a FWHM $= 95.7$~meV. c) Room temperature RIXS spectrum of \Na{} measured at $\Gamma$ for 15 min. d) Transient light-induced modulation of the Bi (111) Bragg peak intensity after arrival of a 1900 nm laser pulse with $F = 97$ mJ cm$^{-2}$.}
    \label{fig:beamline}
\end{figure}

Despite extensive efforts to discover materials exhibiting quantum spin liquid behavior under equilibrium conditions, such discoveries have remained elusive. The process of renormalizing magnetic exchange parameters through photoassisted virtual hopping pathways offers a promising opportunity to identify potential topological excitations within the quantum spin liquid. In this regard, the use of tr-RIXS experimental setup at the XFEL facility emerges as the primary spectroscopic tool for monitoring the emergence of non-equilibrium quantum phases, with particular attention to intensity modulations as a function of momentum, which provide critical insights into these phenomena \cite{Dean2016,tr_RIXS_Review,Mitrano2020,PNAS_327,trRIXS_theory_2,Bernina}.

The tr-RIXS experiment was conducted at the X-ray Scattering and Spectroscopy (XSS) hutch of the PAL-XFEL \cite{park2016}. Self-seeded hard x-ray pulses were generated with the electron bunch charge of 170 pC and the undulator $K$ parameter of 1.87 with a period of 28 mm \cite{nam2021}. A narrowband seed pulse was filtered after eight 5 m-long undulators by forward Bragg diffraction (FBD) of the $[224]$ peak through a 100 \textmu{}m-thick diamond [100] crystal \cite{min2019}. The seed pulse was then directed by a bandpass filter with a delay of tens of femtoseconds, and overlapped in time with the detoured electron bunch by a magnetic chicane to amplify the narrow seed intensity in the following undulators \cite{lee2015}. The incident XFEL beam was assessed by diagnostic tools placed along the beamline from upstream to downstream \cite{choi2023}. To calibrate the incident photon energy, we measured the x-ray absorption spectrum of IrO$_{2}$ powder sample in total fluorescence yield at room temperature, as shown in Figure \ref{fig:beamline} a).

For the RIXS setup, a spherically bent analyzer (diameter = 100 mm, radius of curvature = 1000 mm) with Si(844) diced crystals (2 x 2 mm$^2$) was positioned at $2\theta \!=\! 90^{\circ}$. A two-dimensional (2D) JUNGFRAU detector (pixel size = 75 \textmu{}m) was aligned to the Rowland circle for the Bragg reflection of $E_{i} = 11214$ eV ($\theta_{B} \!=\! 86^{\circ}$) in the vertical diffraction plane. The detector’s active area was shielded with lead tape except for an opening to the crystal analyzer to minimize background. The Bragg angle of the Si(844) crystal analyzer was slightly increased such that the elastic peak is close to the highest energy within the 1.6 eV energy dispersion window of the Si(844) dice. The output from the JUNGFRAU was converted from pixel to energy by applying a calibration factor of $epp = 31.87$~meV pixel$^{-1}$. An avalanche photodiode (APD) was installed on the same side of the Si(844) crystal analyzer to monitor x-ray fluorescence from the sample. A helium-flowing environment was made along the x-ray flight paths. The self-seeded pulses at a repetition rate of 60 Hz were then focused to a spot size of 20 \textmu{}m in full width at half maximum (FWHM) at the sample position by a set of beryllium compound refractive lenses (CRLs). Unless specified, the average x-ray pulse energy was 15 \textmu{}J after the CRLs. To calibrate the energy resolution of the spectrometer, we measured elastic scattering signal from a Scotch tape while tuning a double-crystal monochromator (DCM) from Si(111) to Si(333) Bragg diffraction with the electron bunch energy and the FBD of diamond [100] crystal kept at their optimal condition for the Ir $L_3$ edge, $E_{i} = 11214$ eV. Figure \ref{fig:beamline} b) shows the reference elastic signal for the spectrometer with an optimal resolution of $\Delta E = 95.7$~meV, extracted from a fit to a Pseudo-Voigt line shape. The sample environment was set by a nitrogen-flow cryostat with a base temperature of $T \!=\! 100$~K. Figure \ref{fig:beamline} c) shows an energy loss spectrum ($E_{loss} = E_{i} - E_{f}$) at $\Gamma$ for a \Na{} single crystal with optimal resolution measured at room temperature. To prevent radiation damage on the sample at resonance, the x-ray was defocused to a spot size of 146 \textmu{}m, which degraded the resolution to $\Delta E \approx 120$~meV.

1900 nm laser pulses (100 fs in FWHM) were synchronized to the half repetition rate of the XFEL pulses, i.e., 30 Hz, and propagated at the incident angle of 10 \textdegree{} with respect to the XFEL. Data acquisition by the JUNGFRAU detector was synchronized to the XFEL repetition rate and tagged (laser on, laser off) for data analysis. The laser was focused to a spot size of 340 \textmu{}m in FWHM at the sample interaction point, and its fluence was set to 97 mJ cm$^{-2}$. A spatiotemporal overlap between the x-ray and 1900 nm laser pulses was determined by tracking the transient modulation of the Bismuth (111) Bragg peak intensity \cite{kang2017}. The decay at $\Delta t = 0$~ps and subsequent oscillations of the Bragg peak intensity due to photoexcited phonon dynamics observed in Figure \ref{fig:beamline}(d) confirm the stable arrival time and synchronization of the pair of XFEL and laser pulses without the need for timing-jitter corrections. The overall time resolution, i.e., instrument response function, of the tr-RIXS setup was determined to be 167 fs in FWHM.

\section{Results and discussion}

\begin{figure}
    \centering
    \includegraphics[width=1\textwidth,trim=0.5cm 0cm 0cm 0cm,clip]{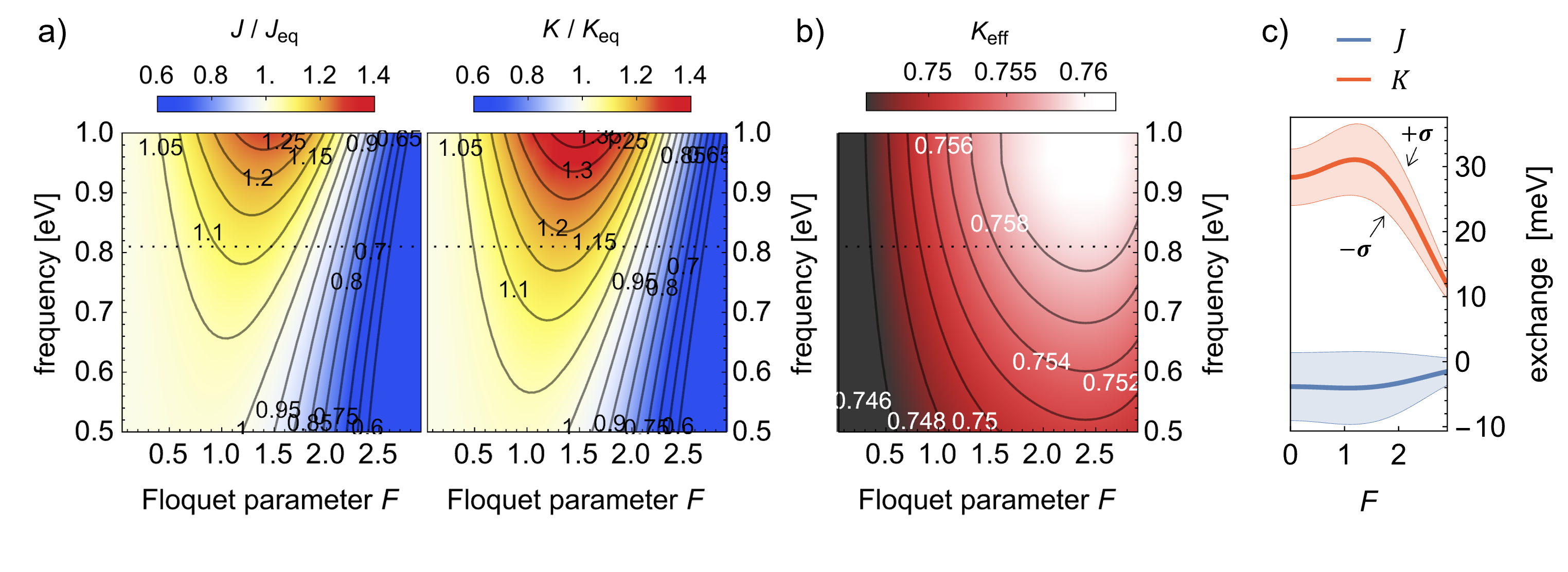}
    \caption{Floquet Control of Exchange Parameters in H\textsubscript{3}LiIr\textsubscript{2}O\textsubscript{6}. (a) Light-induced modification of Heisenberg ($J (A,\omega)$) and Kitaev ($K(A,\omega)$) exchange couplings, with respect to their values in equilibrium ($J_{\rm eq}, K_{\rm eq}$), as a function of the Floquet parameter $F$ and pump frequency. (b) Normalized Floquet Kitaev exchange parameter $K_{eff} = K / [|J| + |K| + |\Gamma|]$. (c) Mean and standard deviation of the Floquet Heisenberg and Kitaev exchange parameters for $\Omega = 0.65 {\rm eV}$, averaged over variations of the electronic parameters (inter-orbital hopping and Coulomb interactions).}
    \label{fig:calculations}
\end{figure}

To capture the pump-induced modification of Kitaev and Heisenberg exchange couplings, we study a minimal model of two neighboring Ir ions coupled via ligand-mediated electron tunneling processes. A single hole resides on average in the $t_{2g}$ orbital manifold per Ir site, which splits into $j_{\rm eff} = 1/2$ and $j_{\rm eff} = 3/2$ states via strong spin-orbit coupling $\lambda$. In the large cubic crystal field limit, a combination of strong local Coulomb ($U$) and Hund's ($J_H$) interactions open a gap for charge excitations. The Kanamori-type Hamiltonian $H_U$ is given by
\begin{align}
    \hat{H}_0 &= U \sum_{i\alpha} \ND{i\alpha\uparrow} \ND{i\alpha\downarrow} ~+ \sum_{i\sigma\sigma',\alpha<\beta} (U' - \delta_{\sigma\sigma'} J_H) \ND{i\alpha\sigma'} \ND{i\beta\sigma} \notag\\
    &+ J_H \sum_{i,\alpha \neq \beta} \left( \CD{i\alpha\uparrow} \CD{i\alpha\downarrow} \C{i\beta\downarrow} \C{i\beta\uparrow} - \CD{i\alpha\uparrow} \C{i\alpha\downarrow} \CD{i\beta\downarrow} \C{i\beta\uparrow} \right) ~+~ \frac{\lambda}{2} \sum_i \mathbf{c}^\dag_i (\mathbf{L}\cdot \mathbf{S}) \mathbf{c}_i
\end{align}
with $i$ summing over the Ir sites, and $\alpha$, $\beta$ indexing the t\textsubscript{2g} orbitals of Ir. We assume $C_3$ rotation symmetry, which neglects a weak additional crystal field splitting.

Hopping between Ir sites is modeled by four Slater-Koster parameters  $t_{xz,xz}$,$t_{xz,yz}$, $t_{xy,xy}$ and $t_{xz,xy}$. Electrons are coupled to light via the Peierls substitution $\CD{i\alpha} \C{j\alpha'} \rightarrow e^{i \frac{e}{\hbar} \mathbf{r}_{ij} \cdot \mathbf{A}(t)} \CD{i\alpha} \C{j\alpha'}$, where $\mathbf{r}_{ij}$ is a bond vector and $\mathbf{A}(t)$ is the vector potential. For a periodic drive with frequency $\omega$, this defines a (dimensionless) Floquet parameter $\mathbf{r}_{ij} \cdot \mathbf{A}(t) = F \cos(\omega t)$ where $F = a_0 e \mathcal{E} / (\hbar \Omega)$ is defined via the peak electric field strength $\mathcal{E}$ and Ir-Ir distance $a_0$. The hopping Hamiltonian between two sites $i$ and $j$ along a $z$ bond then reads
\begin{align}
    &\hat{H}'_{ij}(t) = \sum_{\sigma} e^{i F \cos(\omega t)} \left[ \CD{id_{xz}\sigma}~ \CD{id_{yz}\sigma}~ \CD{id_{xy}\sigma} \right] \hspace{-0.1cm} \cdot \hspace{-0.1cm}  \left[\begin{array}{ccc} t_{xz,xz} & t_{xz,yz} & t_{xz,xy} \\
    t_{xz,yz} & t_{xz,xz} & t_{xz,xy} \\
    t_{xz,xy} & t_{xz,xy} & t_{xy,xy} \end{array}\right] \hspace{-0.1cm} \cdot \left[ \CD{jd_{xz}\sigma}~ \CD{jd_{yz}\sigma}~ \CD{jd_{xy}\sigma} \right]
\end{align}
following the model discussed in Ref. \cite{PhysRevLett.121.247202,PhysRevLett.112.077204,PhysRevB.88.035107}.

If the pump field with frequency $\omega$ is off-resonant from $d-d$ excitations, then charge excitations remain largely suppressed provided that the pulse duration is sufficiently short. In this case, Floquet modifications of spin exchange processes can be readily computed by simultaneously integrating out charge excitations (two-hole Ir states) \textit{and} photon absorption processes via a Schrieffer-Wolff transformation \cite{Claassen_RuCl3,PhysRevB.103.L100408,Kumar2022}. The resulting spin Hamiltonian for a $\gamma$-direction bond takes the form
\begin{equation}
    \hat{H}^{(\gamma)}_{ij} = \sum_{\left<ij\right>_\gamma, \alpha \beta (\gamma)} J \hat{\mathbf{S}}_i\cdot\hat{\mathbf{S}}_j + K \hat{S}^{\gamma}_{i} \hat{S}^{\gamma}_{j} + \Gamma (\hat{S}^{\alpha}_i \hat{S}^{\beta}_j + \hat{S}^{\beta}_i \hat{S}^{\alpha}_j)
\end{equation}
with Kitaev $K$, Heisenberg $J$ and off-diagonal interactions $\Gamma$ interactions. We neglect weak anisotropic interactions $\Gamma'$ of the form $\hat{S}^\alpha_i \hat{S}^\gamma_j$ \cite{PhysRevB.93.214431}. Within our Floquet-Kitaev model, all exchange interactions become functions of the driving field amplitude $A$ and frequency $\omega$ ($K(A,\omega)$, Heisenberg $J(A,\omega)$, $\Gamma$). Physically, this dependence arises from effective photon-dressed exchange processes, whereby a hole virtually tunnels to a neighboring Ir site while absorbing a photon to mitigate its intermediate-state energy penalty due to another hole on the same site; the second hole can subsequently tunnel back and re-emit a photon \cite{Claassen_RuCl3,PhysRevB.103.L100408,Kumar2022}. Figure \ref{fig:calculations}a) shows the calculated light-induced modification of $K(A,\omega)$, Heisenberg $J(A,\omega)$ with respect to the equilibrium values for $U = 2.2$~eV, $J_H = 0.3$~eV, $\lambda = 0.540$~eV. We fix inter-Ir hopping values $t_{xz,xz} = 0.050$~eV,$t_{xz,yz} = 0.440$~eV following Ref. \cite{delaTorre2023}, and choose the remaining hopping parameters $t_{xy,xy} = 0.08$~eV and $t_{xz,xy} = 0.04$~eV, consistent with \cite{PhysRevLett.121.247202} and an expected equilibrium Kitaev exchange interaction $\sim 30$~meV, and checking sensitivity to these choices below. The corresponding normalized Floquet Kitaev parameter ($K_{eff} = K / [|J| + |K| + |\Gamma|]$) is shown in Figure \ref{fig:calculations}b). While across the range of Floquet parameters and pump wavelengths considered in our calculations, we observed the enhancement and suppression of the exchange interactions, $K_{eff}$ shows a small increase for all parameters. This is suggestive that Floquet engineering can bring \hlio{} proximal to the Kitaev limit. However, we note that this result is extremely sensitive to changes of $U$ and the inter-orbital hopping parameters $t_{xy,xy}$ and $t_{xz,xy}$. To account for this variation, in Figure \ref{fig:calculations} c) we show the mean and standard deviation of $J(A,\omega)$ and $K(A,\omega)$ computed for a range of parameters $U \in [1.9,2.4]$~eV, $t_{xy,xy} \in [-0.14,0.14]$~eV and $t_{xz,xy} \in [-0.1,0.1]$~eV.

\begin{figure}
    \centering
    \includegraphics[width=1\linewidth]{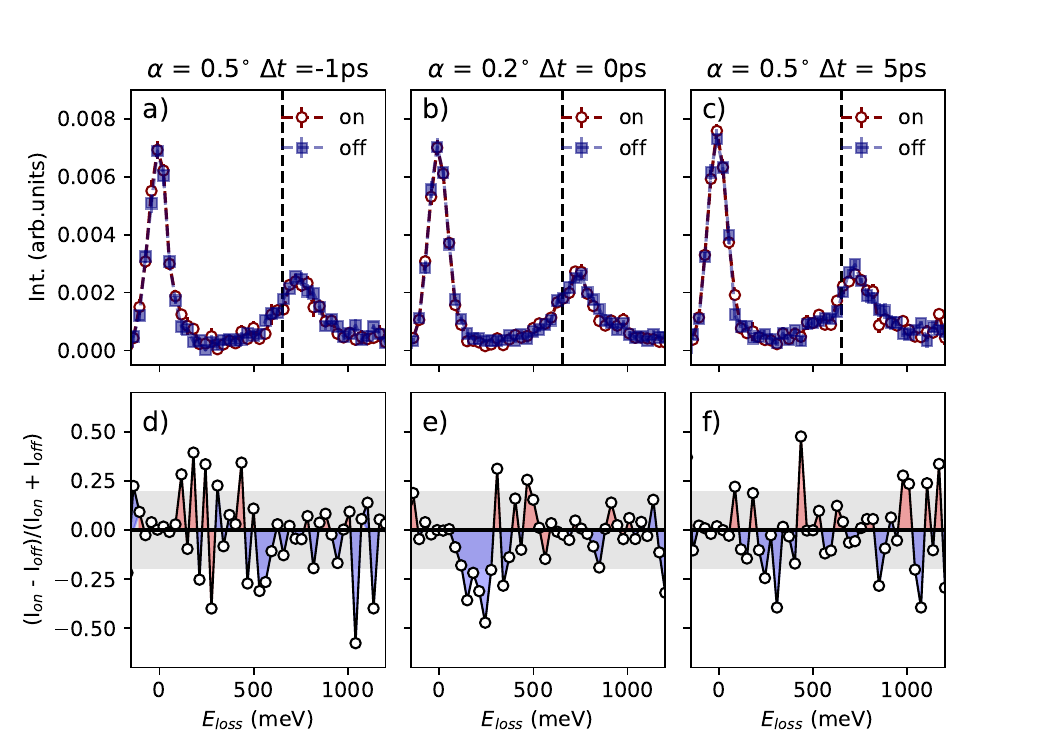}
    \caption{Floquet engineering of the magnetic excitations in \hlio{}. a)-c) RIXS spectrum at resonance and $T = 100$~K with (on, red circular markers) and without (off, blue square markers) a $1900$~nm laser pulse at various time delays: a), $\Delta t = -1$~ps, the average of 5 individual scans of 12 min exposure for each condition b), $\Delta t = 0$~ps, the average of 10 individual scans, and c), $\Delta t = 5$~ps, the average of 4 individual scans. The vertical dashed line signal encodes the energy of the laser pump. $\alpha$ is the angle of incidence of the x-ray. d)-e) Relative change of the RIXS spectrum due to the laser pulse at d), $\Delta t = -1$~ps, e), $\Delta t = 0$~ps, and, f), $\Delta t = 5$~ps scans. The gray bar illustrates the mean standard deviation. Red and blue shading indicates increased or decreased intensity due to the laser excitation.}
    \label{fig:data}
\end{figure}

Although our model shows a small dependence of $K_{eff}$ with $\omega$ (Figure \ref{fig:calculations}), we set the pump laser 1900 nm such as the drive photon energy is proximal but non-resonant with the Ir intra-$t_{2g}$ excitations to minimize heating effects,u \cite{delaTorre2023}. The laser fluence is set to $F = 97$ mJ/cm$^{2}$, which corresponds to a peak electric field $E = 5.4$~GV/m, two orders of magnitude larger than the driving field needed to generate Floquet states in graphene \cite{merboldt2024observationfloquetstatesgraphene,choi2024directobservationfloquetblochstates}. The corresponding Floquet parameter $F = aeE/\hbar\Omega = 2.56$, with $a = 3.1$~\AA, the interatomic Ir-Ir distance, $e$ the electron charge and $\Omega$ the frequency of the driving laser, is comparable to that used to engineer optical nonlinearities in a 2D van der Waal material \cite{Shan2021}. 

The main experimental results of our work are summarized in Figure \ref{fig:data}. In Figure \ref{fig:data} a)-c), we show the RIXS spectra at grazing incidence ($\alpha = 0.5^{\circ}$ or $\alpha = 0.2^{\circ}$) from \hlio{} with and without the incident 1900 nm circularly polarized laser pulse of $F = 97$ mJ/cm$^2$ at three different time delays: a), $\Delta t = -1$~ps, b), $\Delta t = 0$~ps, and c), $\Delta t = 5$~ps. To evaluate the effect of the circular pump at 1900 nm on the magnetic and charge excitations of \hlio{}, we plot the relative change, I$_{on}$-I$_{off}$/(I$_{on}$+I$_{off}$), in Figure \ref{fig:data} d)-f). We only observe changes of the RIXS intensity above the mean standard deviation during the sample illumination at $\Delta t = 0$~ps [Figure \ref{fig:data} e)]. On the contrary, at later, $\Delta t = 5$~ps [Figure \ref{fig:data} f)], and earlier time delays, $\Delta t = -1$~ps [Figure \ref{fig:data} d)], no significative intensity change is observed. The suppression of the RIXS spectral weight at $\Delta t = 0$~ps occurs in the $E_{loss} \in [50,300]$~meV range, which encompasses the tail of the broad magnetic continuum in \hlio{} associated with the presence of fractionalized excitations \cite{delaTorre2023}. Moreover, the lack of significative change to the intra-$t_{2g}$ excitations $E_{loss} \in [500,1200]$~meV intensity indicates the absence of heating and decoherence at $\Delta t = 0$~ps despite the large peak $E$-fields. Our data suggest that Floquet engineering by a circularly polarized pulse in \hlio{} leads to an increase in the coherence of the magnetic excitations, possibly due to \hlio{} approaching a ferromagnetic ground state.

\begin{figure}
    \centering
    \includegraphics[width=1\linewidth]{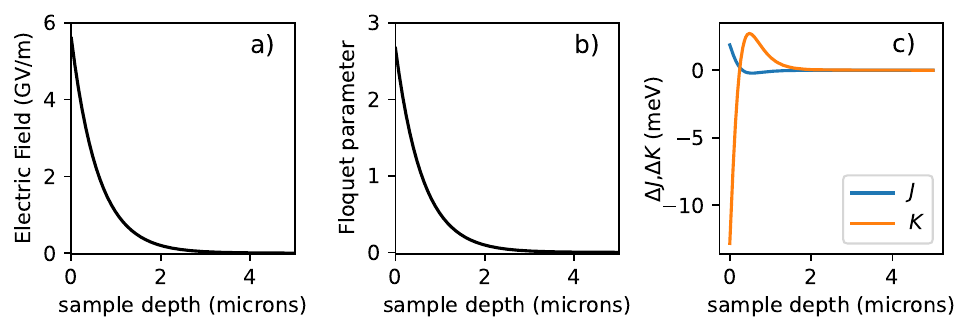}
    \caption{Effects of the penetration depth mismatch. Evolution of a) the Electric field, $E,$, b) the Floquet parameter,$F$ and c) Floquet-induced modification of the exchange interactions, $\Delta J = J(A,\omega) - J_{eq}; \Delta K = K(A,\omega) - K_{eq}$ as a function of the sample depth. }
    \label{fig:penetration_mismatch}
\end{figure}

These results contrast with our Floquet calculation (Fig. \ref{fig:calculations}). Given our experimental conditions with $F = 2.56$ should lead to a reduction of $K = 10$~meV and a full suppression of $J$ thus bringing \hlio{} into the KQSL limit. However, the penetration depth mismatch between the pump laser and the hard X-ray probe prevent us from probing the large $F$ regime. Figure \ref{fig:penetration_mismatch}a)-b) shows the evolution of $E$ and $F$ within the sample depth assuming an absorption coefficient $\alpha = 300$~nm \cite{alpha-absortion_coefficeint}. After the first $2 \mu$m, $F$ is negligible, and so are the effects of the laser electric field on the exchange interactions of \hlio{} (Fig. \ref{fig:penetration_mismatch}c). The average Floquet parameter over the sample estimated thickness ($d = 5 \mu$m) $\bar{F} = 0.128$ would lead to small changes of $\Delta K/K_{eq} = 0.12 \%$ and $\Delta J/J_{eq} \approx -0.7 \%$, below the resolution of our experiment. The dependence of $J$ and $K$ on model parameters should also be considered in the low $F$ regime, making calculating the expected changes to the exchange interactions difficult.

\section{Conclusions and Outlook}

Altogether, our theoretical calculations and experimental results establish Floquet engineering as a new approach to tune the magnetic Hamiltonian of Kitaev magnets. However, the intrinsic complexity of time-resolved hard X-ray experiments in light engineered quantum magnets, namely exchange frustration, structural disorder, decoherence, and penetration depth mismatch, hinders us from conclusively demonstrating the realization of the light-induced KQSL. To unequivocally demonstrate the light-matter engineering of long-range entangled magnetic ground states in Kitaev magnets, a systematic approach combining improved material synthesis, pump-probe-specific sample fabrication and methodology, theory, and experiment development is needed. In the context of \textit{Kitaev magnets for out of equilibrium}, a possible avenue is to consider other new generation iridium-based Kitaev magnets beyond \hlio{} \cite{molecules27030871}. Recently, D$_3$LiIr$_2$O$_6$ \cite{halloran2024} has emerged as an alternative since the heavier deuterium atoms will lead to fewer stacking faults in the structure than in \hlio{}. Alternatively, the use of exfoliated flakes or thin films of iridium-based Kitaev magnets could minimize the penetration depth mismatch between the laser pump and X-ray probe. On the other hand, pump and probe penetration depths are comparable in Kitaev materials with transition metal oxides with $L$ edges in the soft X-ray regime. We highlight earth-abundant cobalt-based Kitaev material, such as Na$_2$Co$_2$TeO$_6$, as a platform for future experiments \cite{dufault_introducing_2023}. The results presented serve as stepping stone towards achiving the elusive Kitaev limit and will motivate future tr-RIXS experiments in Kitaev magnets as well as presenting the need to develop a formal theory of fractionalized excitations in Floquet KQSLs and their signatures in tr-RIXS.

\begin{backmatter}
\bmsection{Funding}

The experiment was performed using the FXS instrument at the PAL-XFEL (Proposal No. 2024-1st-XSS-013) funded by the Ministry of Science and ICT of Korea (RS-2022-00164805). The work at Boston College was funded by the U.S. Department of Energy, Office of Basic Energy Sciences, Division of Physical Behavior of Materials under award number DE-SC0023124. M.~C. acknowledges support from the U.S. Department of Energy, Office of Basic Energy Sciences, under Award No. DE-SC0024494.

\bmsection{Acknowledgment}

Sample characterization by XRD was performed at the Canadian Light Source, a national research facility of the University of Saskatchewan, which is supported by the Canada Foundation for Innovation (CFI), the Natural Sciences and Engineering Research Council (NSERC), the Canadian Institutes of Health Research (CIHR), the Government of Saskatchewan, and the University of Saskatchewan.  Experiments at 3A beamline of PLS-II were supported in part by MIST. The use of the Advanced Photon Source at the Argonne National Laboratory was supported by the US Department of Energy (Contract DE-AC02-06CH11357). The authors thank all staff members of the PAL-XFEL for supporting the tr-RIXS experiment. The authors thank the Global Science Experimental Data Hub Center (GSDC) at the Korea Institute of Science and Technology Information (KISTI) for providing computing resources and technical support. A.d.l.T. acknowledges helpful conversations with Kemp Plumb, Hui-Yuan Daniel Chen , and Greg Fiete. 

\bmsection{Disclosures}

The authors declare no conflicts of interest.

\bmsection{Data Availability Statement}

Data underlying the results presented in this paper may be obtained from the authors upon reasonable request.



\end{backmatter}

\end{document}